# Social Networks Analysis to Retrieve Critical Comments on Online Platforms

Shova Bhandari
George Mason University
sbhand7@gmu.edu

Rini Raju
George Mason University
rriniraj@gmu.edu

[1-2] ***Abstract*—Social networks are rich source of data to analyze user habits in all aspects of life. User's behaviour is decisive component of a health system in various countries. Promoting good behaviour can improve the public health significantly. In this work, we develop a new model for social network analysisby using text analysis approach. We define each user reaction to global pandemic with analysing his online behaviour. Clustering a group of online users with similar habits, help to find how virus spread in different societies. Promoting the healthy life style in the high risk online users of social media have significant effecton public health and reducing the effect of global pandemic.In this work, we introduce a new approach to clustering habits based on user activities on social media in the time of pandemic and recommend a machine learning model to promote health in the online platforms.**

## I. Introduction

Corona viruses are a large family of viruses that in humans are known to cause respiratory infections ranging from the common cold to more severe diseases such as Middle East Respiratory Syndrome (MERS) and Severe Acute Respira-tory Syndrome (SARS) [1]. The novel coronavirus of 2019 (COVID-19) is a contagious virus first identified in Wuhan, China. COVID-19 is now a pandemic affecting many countries globally [1].

The World Health Organization (WHO) declared a public health emergency on January 30, 2020, and officially named the virus COVID-19 on February 12, 2020. The WHO declared COVID-19 a pandemic on March 11, 2020. More than 67 million people are infected, and 1.54 million people have died due to this virus globally [2]. The United States of America is one of the hardest hit areas in the world where more than 14.8 million infected, and more than 248K people died as of December 6, 2020 [2].

COVID-19 spreads from person to person through respiratory droplets. The most common symptoms of this virus are sore throat, dry cough, fever, loss of smell, and difficulty breathing. Even though COVID-19 affects humans of all age groups, adults over 60 and individuals with various health issues (i.e., pre-existing conditions) are affected the most.

Aside from physical symptoms, the COVID-19 pandemic has affected living conditions of many individuals, as well as business and commerce across the globe. Most businesses are going bankrupt due to the sudden downfall of the resources, supply chain, and economy. COVID-19 has spread rapidly across the globe, the world is fighting to defeat this pandemic. To defeat this pandemic many countries underwent lockdown to slow the spread of infection. To reduce the negative overall impacts on the global economy, many countries are developing a vaccine to fight this virus.

Dealing with the unforeseen challenges caused by the COVID-19 pandemic has taken a significant toll on the global economy and population. Assessing the impact of the COVID-19 crisis on society, economy and other vulnerable sectors/groups is fundamental to inform and tailor government responses. As well as creating partnerships to recover from the crisis and ensure that no one is left behind in this effort. However, even though COVID-19 presents significant issues there are some groups of individuals that call it a hoax or trying to use propaganda to legitimize the dangers associated with this virus. Therefore, the use of facts and numerical evidenceis required to dispel these falsehoods [3] [4] [5] [6]

Therefore, it is important to predict COVID-19 cases, cases that are likely to be hospitalized this fall. If the cases are accurately predicted, hospitals can establish a separate COVID-19 unit to treat and isolate the COVID-19 patient reducing the anticipated strain on the healthcare system [7] [6], [8]– [13]. Furthermore, even though COVID-19 presents significantissues there are some groups of individuals that call it a hoax or trying to use propaganda to legitimize the dangers associated with this virus. People are picking up COVID-19 conspiracy theories from social media, especially twitter. The use of facts and numerical evidence is required to dispel these falsehoods. One way to determine overall impact is to evaluate how misinformation (or conspiracy theories) affect infection, hospitalization, and death rates for specific regions/locations. Expert fake account detection methods [14] [15] can stopspread of misinformation.

COVID-19 research so far has been based on the survey of 100-1000 individuals. We did not find any literature paper based on circulation of the covid-19 misinformation on social media. Social media have far more reach than the traditional news in current days and people blindly follow, retweet and like the misinformed news on social media [16] without analyzing the truthfulness of the news. Therefore, we decided to analyze twitter data for misinformation and compare itCOVID-19 daily cases, hospitalization, and death rates for specific regions/locations can be used in transfer learning

```
In [49]:  df=pd.read_csv("all-states-history.csv")
          print (df.shape)

          (15409, 42)
```

Fig. 1. Figure 1

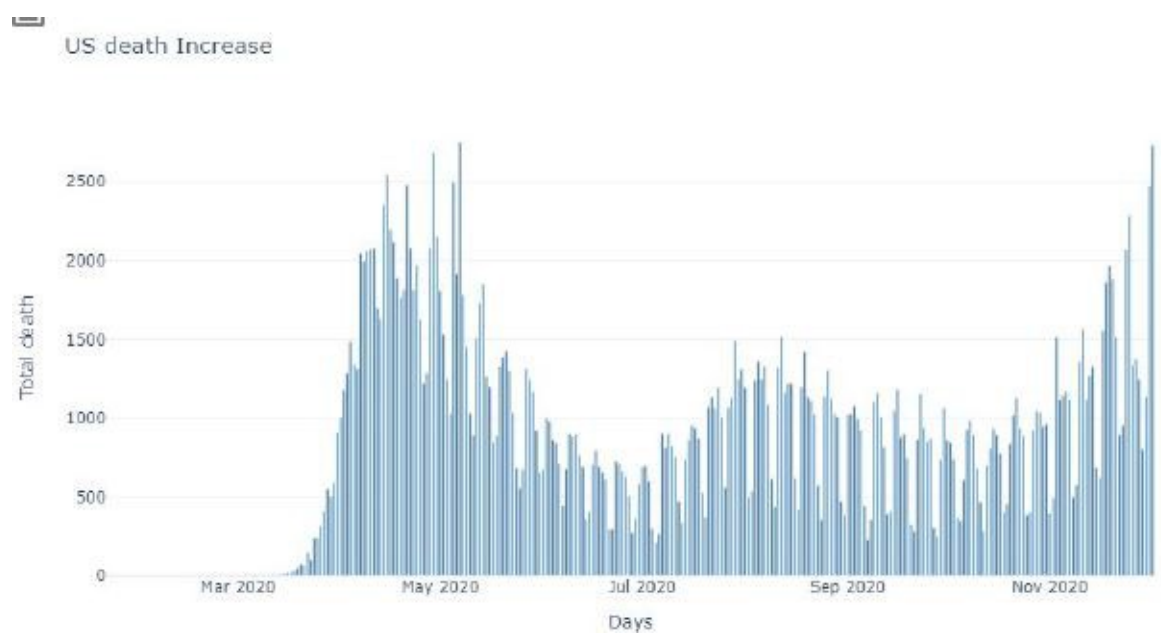

Fig. 2.

methods to analyse data [17] [18] [19].

To understand the seriousness of the pandemic and its effects on individuals, sectors, countries, and the entire world, it is important to present the facts rather than an unproven hypothesis. Specifically, to individuals using propaganda to dispel the severity of COVID-19. Metrics and visualizations provide a graphical representation (or story) as to the seriousness of the problem. Therefore, major factors must be taken into consideration to provide the indisputable facts about the seriousness of COVID-19. A few examples of which are:

## II. Dataset

This paper is based on COVID-19 data [20]. This dataset was acquired from COVID-19 Tracking project and NY times. COVID-19 adat which is used to draw a visualization on different questions like the number of confirmed cases, total number of deaths, number of recovery and so on for worldwide cases. This data set has 156292 records and 8 fields.

Another data set used is all-states-history.csv extracted from COVIDtracking.com. This data contains the daily positive, hospitalized, deaths and recovered cases for US States. This data has 15409 records (rows) and 42 columns. see the the following figures 1 3 2.

The above visualization displays the COVID-19 cases in the USA. The graph shows the daily cumulative count of confirmed cases, daily cumulative count of deaths and daily count of new confirmed cases in the USA. The X-axis indicates the date of observation.

We also extracted tweet data related to COVID-19 misinformation using Twitter API.

## III. Methods

For desired results, COVID-19 related dataset from Kaggle and COVID tracking.com is processed and analyzed. Cases, deaths, and hospitalization numbers based on these datasets reflect cumulative totals since January 22, 2020 until December 2, 2020. Based on the frequency of how Kaggle and COVID-19 tracking website updates these data, they may not indicate the exact numbers of daily, hospitalization, and death cases as reported by the state and local government organizations or the news media. This paper uses daily new cases, hospitalization, and COVID-19 related death data sets. See the the following figures 5 6 7.

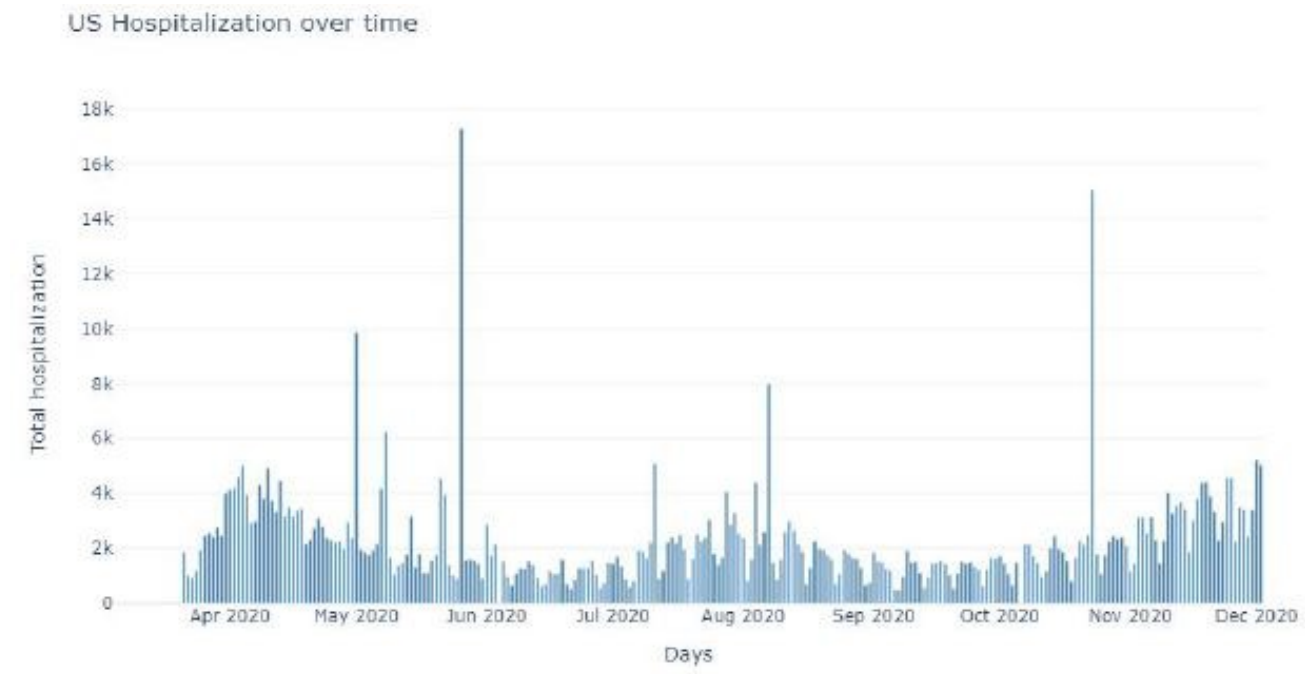

Fig. 3.

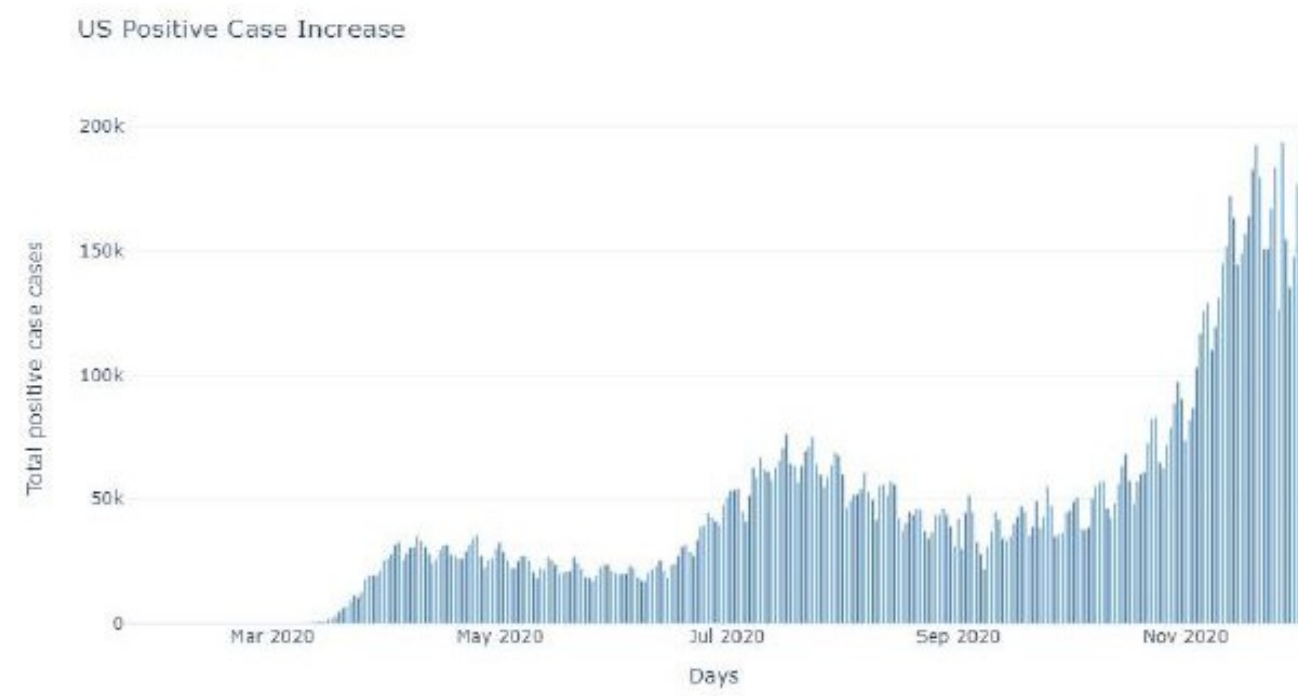

Fig. 4. Figure 1

For misinformation, controversy and hoax, Twitter data is collected using Twitter API. This data set has 17560 records and 7 fields.

Data is analyzed using Python, Natural Language Processing (NLP) and Tableau. Implementation of these analysis

```
# pass these authorization details to tweepy
api = tw.API(auth, wait_on_rate_limit=True)
# test authentication
try:
    api.verify_credentials()
    print("Authentication OK")
except:
    print("Error during authentication")
# Extracting Specific Tweets from Twitter
search_words = 'plandemic'
new_search = search_words +" -filter:retweets"
```

Fig. 5.

```
#Add time your file was created to discr
filename = 'Covid19_tweet_Worldwide.csv'
```

Fig. 6.

```
# Open/Create a file to append data
with open (filename, 'a', newline='') as csvFile:
   csvWriter = csv.writer(csvFile)
   for tweet in tw.Cursor(api.search,q=new_search, count
                              lang="en",
                              tweet_mode= 'extended',
                              since='2020-11-01',
                              until ='2020-12-03').items():
        #tweets_encoded = tweet.text.encode('utf-8')
        #tweets_decoded = tweets_encoded.decode('utf-8')
        tweetCountTest += 1
        print(tweetCountTest)
        print (tweet.created_at, tweet.full_text)
       # if tweet.coordinates or tweet.geo:
        csvWriter.writerow([tweet.created_at, tweet.full_
```

Fig. 7.

processes includes cleaning/standardizing and pre-processing of the dataset to get rid of any duplicate values or outliers. See the the following figures 8 9 10.

Place field of Twitter data was in json format. Therefore, this field was split into type, coordinate, city, state, country code and country.

This paper then studies and analyzes the cleaned data for better representation of the result. With the transformed data, we performed an exploratory and predictive analysis.

Based on the major factors outlined in Section I the following questions were derived:

1. How did pre-emptive measures impact overall transmission/infection rates?

2. How did controversy impact preemptive measures and transmission infection rates?

3. Compare the number of retweeted COVID-19 misinformation/conspiracies in relation to hotspots (i.e., States with the highest number of infections, hospitalizations, and or deaths).

This paper recognizes that the nature of this topic is ideological in some regard, research on ideological assumptions must also be taken into consideration when extrapolating answers to these questions. Our initial approach as previously explained is to gather numerical based data for the purposes of providing a quantitative analysis. After we completed the analysis, we took the data and developed visualizations using Tableau and Python. See the the following figures 11, 12.

Using various libraries, we then imported the related information and developed visualizations of the results for a better understanding. We then used a forecast and time series model to forecast future COVID-19 case and death trends. Our team also incorporated a secondary approach to this research. We performed a social media Sentiment Analysis to explore COVID-19 tweets that potentially impacted how individuals interpreted the seriousness of COVID-19. This methodology (or behavioral analysis) allowed us to identify a specific user

```
data.head()
```

| | Date_Time | text | username | user_location | retweet_count | favourite_count | Place |
|---|---|---|---|---|---|---|---|
| 0 | 12/2/2020 23:51 | b'@realDonaldTrump Wait! Didn\xe2\x80\x99t you say the Samething about COVID19? Hoax/FakeNews!' | b'DavidTashawn' | b'Mexico, MO' | 0 | 0 | NaN |
| 1 | 12/2/2020 23:45 | b'Dr. Henry on people who think #COVID19 is a hoax: "Might I suggest that you don\'t follow Twit... | b'katslepian' | b'Surrey, British Columbia' | 1 | 7 | NaN |
| 2 | 12/2/2020 23:32 | b'#COVIDIOTS\xf0\x9f\x92\xa5You will soon understand your ignorance protected a HOAX that was me... | b'AndreeaSaint' | b'Canada' | 0 | 2 | NaN |
| 3 | 12/2/2020 23:27 | b"@stephanieinnes @azcentral I can't believe there are still people who think #COVID\xe3\x83\xbc... | b'roanneaz63' | b'Arizona' | 0 | 9 | NaN |
| 4 | 12/2/2020 23:21 | b'Many Republicans predicted that as soon as Democrats won in November the Covid19 virus would d... | b'mkevane' | b'Tweets are personal.' | 1 | 6 | NaN |

```
print('Dataset size:',data.shape)
print('Columns are:',data.columns)

Dataset size: (17560, 7)
Columns are: Index(['Date_Time', 'text', 'username', 'user_location', 'retweet_count',
       'favourite_count', 'Place'],
```

Fig. 8.

```
data.info()

<class 'pandas.core.frame.DataFrame'>
RangeIndex: 17560 entries, 0 to 17559
Data columns (total 7 columns):
 #   Column           Non-Null Count  Dtype
---  ------           --------------  -----
 0   Date_Time        17560 non-null  object
 1   text             17560 non-null  object
 2   username         17560 non-null  object
 3   user_location    17560 non-null  object
 4   retweet_count    17560 non-null  int64
 5   favourite_count  17560 non-null  int64
 6   Place            277 non-null    object
```

Fig. 9.

group and keywords used when commenting on COVID-19 information. It helps to identify the number of people who believe on the controversy related to COVID such as:

COVID-19 is man made in the laboratory?

It is not real and is a hoax?

5G technology is responsible for the global pandemic?

As with our initial approach, the Sentiment Analysis leveraged both Python and R/R Studio as programming languages and visualization tools. We developed a custom code using Python that used an application programming interface (API) to interact with Twitter. The Python code contains keys and tokens created within the Twitter Development Console to

```
def clean(sentence):
    #make everything lowecase
    sentence = sentence.lower()
    #remove urls
    sentence = re.sub(r'http\S+', " ", sentence)
    # remove mentions
    sentence = re.sub(r'@\w+',' ',sentence)

    # remove hastags
    sentence = re.sub(r'#\w+', ' ', sentence)

    # remove digits
    sentence = re.sub(r'\d+', ' ', sentence)

    # remove html tags
    sentence = re.sub('r<.*?>',' ', sentence)

    #remove stop words
    sentence = sentence.split()
    sentence = " ".join([word for word in sentence if not word in stopwords])
    # remove punctuation
    sentence  = "".join([char for char in sentence if char not in string.punctuation])
    sentence = re.sub('[0-9]+', '', sentence)
     #remove
    sentence = re.sub(r'\bx\d*\b', '', sentence)

    return sentence
```

Fig. 10.

```
#tokenizzation
def tokenization(sentence):
    sentence = re.split('\W+', sentence)
    return sentence

data['Tweet_tokenized'] = data['text'].apply(lambda x: tokenization(x.lower()))
data.head()
```

Fig. 11.

| Field Name | Table | Remote Field Name |
|---|---|---|
| retweet_count | Covid19_tweet_Worldwide.csv | F7 |
| favourite_count | Covid19_tweet_Worldwide.csv | F8 |
| place | Covid19_tweet_Worldwide.csv | F9 |
| ID | | F9 - Split 3 |
| place - Split 7 | | F9 - Split 7 |
| placesplit | | F9 - Split 1 |
| type | | F9 - Split 1 - Split 1 |
| coordinate | | F9 - Split 2 |
| city | | F9 - Split 7 - Split 1 |
| State | | F9 - Split 7 - Split 2 |
| place - Split 8 | | F9 - Split 8 |
| country_codde | | F9 - Split 8 - Split 1 |
| place - Split 9 | | F9 - Split 9 |
| country | | F9 - Split 9 - Split 1 |

Fig. 12.

allow for requests, formatting, and extraction of Twitter as it relates to COVID-19. See the the following figures 13, 14, 15.

In addition to the API logic, our custom code also contains logic for error handling, data cleansing (i.e., removal of special characters and links), classification and parsing of tweets, and calculating positive/negative/neutral Sentiment Analysis percentage ratings.

## IV. Result Analysis

Our team performed several iterations of the Sentiment Analysis using several COVID-19 key words, such as:

COVID-19
COVID-19 hoax
COVID-19 conspiracy
COVID-19 fake news

```
# keys and tokens from the Twitter Dev Console
consumer_key = 'XXXXXXXXXXX'
consumer_secret = 'XXXXXXXXXXX'
access_token = 'XXXXXXXXXXX'
access_token_secret = 'XXXXXXXXXXX'
```

Fig. 13.

```
def clean_tweet(self, tweet):
    '''
    Utility function to clean tweet text by removing links, special characters
    using simple regex statements.
    '''
    return ' '.join(re.sub("(@[A-Za-z0-9]+)|([^0-9A-Za-z \t])|(\w+:\/\/\S+)", " ", tweet).split())
```

Fig. 14.

```
def main():
    # creating object of TwitterClient Class
    api = TwitterClient()
    # calling function to get tweets
    tweets = api.get_tweets(query = 'COVID-19', count = 200)

    # picking positive tweets from tweets
    ptweets = [tweet for tweet in tweets if tweet['sentiment'] == 'positive']
    # percentage of positive tweets
    print("Positive tweets percentage: {} %".format(100*len(ptweets)/len(tweets)))
    # picking negative tweets from tweets
    ntweets = [tweet for tweet in tweets if tweet['sentiment'] == 'negative']
    # percentage of negative tweets
    print("Negative tweets percentage: {} %".format(100*len(ntweets)/len(tweets)))
    # percentage of neutral tweets
    print("Neutral tweets percentage: {} %".format(100*(len(tweets) -(len( ntweets )+len( ptweets)))/len(tweets)))

    # printing first 5 positive tweets
    print("\n\nPositive tweets:")
    for tweet in ptweets[:10]:
        print(tweet['text'])

    # printing first 5 negative tweets
    print("\n\nNegative tweets:")
    for tweet in ntweets[:10]:
        print(tweet['text'])
```

Fig. 15.

COVID-19 spike
COVID-19 hospitalizations
COVID-19 death

Within each of these iterations, the Sentiment Analysis provided three quantifiable metrics for analysis that are Positive, Negative, and Neutral Percentages. The following images reflect the amounts for each of the three metrics for each of the iterations previously identified.

Our analysis shows that the percentages for the search terms of COVID-19 hoax and COVID-19 conspiracy vary between one and three percent. Whereas the COVID-19 fake news search showed significant variation between the previous two search terms.

## V. Discussion

Recognizing that the words hoax, conspiracy, fake news, death, and hospitalizations all have negative connotations associated with them, ideally, the expectation for a Sentiment Analysis associated with these words would reflect a relatively low Positive Percentage. However, the analysis shows that is not the case with the COVID-19 hoax and COVID-19 conspiracy results. Both iterations of these analyses show that there is a Positive Percentage rating of over thirty percent. When compared to the COVID-19 fake news and COVID-19 hospitalization searches the results fell in-line with expectations, reflecting less than fifteen percent for the Positive Percentage rating.

## VI. Analytical conclusion

When performing our analysis we recognized that therewas significant variability in percentage metrics across the search terms identified. As well as, our understanding of whichtweets are classified as positive, negative, or neutral within a given search. In order to conclusively determine an answerto our problem statement significant time has to be spenton reviewing each of the associated tweets to validate the accuracy of the sentiment rating (i.e., positive, negative, and neutral).

In some instances, the body of the Tweet text reflects a negative expression of thought with a positive classification for sentiment. At face value, this has the potential of causing confusion and in some cases misunderstanding/mistrust in the results and analysis. To avoid situations in which the methodology, data collection, data processing, data review, and interpretation it is critical that researchers understand ho w sentiment is classified withinthe context of a specific search.